\documentclass[aps,prb,twocolumn,showpacs]{revtex4}
\usepackage{graphicx}
\usepackage{amssymb}

\newcommand{\CCS}{CdCr$_2$S$_4$}
\newcommand{\HCS}{HgCr$_2$S$_4$}

\begin{document}

\title{Multiferroicity and colossal magneto-capacitance in Cr-thiospinels}

\author{J. Hemberger$^1$, P. Lunkenheimer$^1$, R. Fichtl$^1$,
        S. Weber$^1$, V. Tsurkan$^{1,2}$, and A. Loidl$^1$}

\address{%
$^1$Experimental Physics V, Center for Electronic Correlations and Magnetism, \\
University of Augsburg, D-86135 Augsburg, Germany \\ %
$^2$Institute of Applied Physics, Academy of Sciences of Moldova,
 Chisinau, Republic of Moldova  %
}

%\ead{joachim.hemberger@physik.uni-augsburg.de}

\begin{abstract}

The sulfur based Cr-spinels $R$Cr$_2$S$_4$ with $R$ = Cd and Hg
exhibit the coexistence of ferromagnetic and ferroelectric
properties together with a pronounced magnetocapacitive coupling.
While in \CCS\ purely ferromagnetic order is established, in \HCS\
a bond-frustrated magnetic ground state is realized, which,
however, easily can be driven towards a ferromagnetic
configuration in weak magnetic fields. This paper shall review our
recent investigation for both compounds. Besides the
characterization of the magnetic properties, the complex
dielectric permittivity was studied by means of broadband
dielectric spectroscopy as well as measurements of polarization
hysteresis and pyro-currents. The observed colossal
magneto-capacitive effect at the magnetic transition seems to be
driven by an enormous variation of the relaxation dynamics.

\end{abstract}

% PACS codes here, in the form:
\pacs{75.80.+q; 77.22.Gm}

\maketitle

\section{Introduction}

In recent years multiferroic magnetoelectrics attracted increasing
scientific and technological interest.\cite{fiebig:05jpd} In this
rare class of compounds, ferroelectricity (or at least a weak
ferroelectric component) and (ferro-)magnetism coexist and both
order-parameters are strongly coupled. This coupling of the
dielectric and magnetic properties makes them highly attractive
not only from an academic point of view, but also for potential
applications in microelectronics.
Ferromagnetism is known since ancient times and a prerequisite of
modern technology in many areas. Ferroelectricity has been
identified in 1921 by Valasek\cite{valasek:21pr} in Seignette salt
and the existence of ferroelectricity in an oxide has been
established as late as 1943 in BaTiO$_3$ by von Hippel and
coworkers.\cite{hippel:50rmp} Ferroelectrics play an important
role in thze technological revolution in optics, acoustics, and
capacitor engineering for a variety of applications in electronic
circuitry and modern energy storage. The coexistence of
ferroelectricity and ferromagnetism would constitute another mile
stone for modern electronics and functionalized materials. The
most appealing applications are new types of storage media using
both magnetic {\em and} electric polarization and the possibility
of electrically reading/writing magnetic memory devices (and vice
versa). The first multiferroic material, revealing magnetic {\em
and} electric order to be discovered was nickel-iodine boracite in
1966\cite{ascher:66jap} and nowadays multiferroic materials are in
the focus of interest of recent solid state research.
Prominent examples for such type of materials are the heavy rare
earth manganites like the perovskites TbMnO$_3$ or
DyMnO$_3$,\cite{kimura:03nat,goto:04prl} orthorombic
TbMn$_2$O$_5$,\cite{hur:04nat} the hexagonal YMnO$_3$ and
HoMnO$_3$,\cite{lottermoser:04nat,fiebig:02nat} or the kagom\'e
staircase compound Ni$_3$V$_2$O$_8$.\cite{lawes:05prl} This
contribution summarizes the results on the recently discovered
multiferroic and magneto-capacitive properties of the cubic spinel
systems \CCS\ and \HCS.\cite{hemberger:05nat,weber:06prl} We
report on magnetization, dielectric polarization, as well as
broadband dielectric measurements in these systems and provide
detailed information on the dielectric relaxation dynamics. In
addition, the influence of impurity doping and sample dependence
is addressed.
Spinel compounds are an important class of materials: For example,
spinels are a key constituent in the deep-Earth mantle; and they are
great imposters in gemstone history, often mistaken for rubies, even
in crown jewels. The magnetic properties of spinel-type compounds
have long been known and are in the focus of applied research since
more than 50 years. Ferrites with spinel structure are being used
for many applications in microwave technology and high-frequency
electronics. Spinels have also been in the focus of basic research.
The metal-insulator transition in Fe$_3$O$_4$ was described by
Vervey\cite{vervey:39nat} in 1939 as a charge-ordering transition
but this interpretation remains controversial. Recent reports on
geometrical frustration of the spin and orbital degrees of
freedom,\cite{fritsch:04prl} and the observation of an orbital glass
state\cite{fichtl:05prl} in sulpho spinels, demonstrate the rich and
complex physics, characteristic of these compounds. The richness is
a result of the cooperativity of and the delicate balance between
charge, spin, orbital, and lattice degrees of freedom.

At room temperature \CCS\ and \HCS\ are normal cubic $AB_2$S$_4$
thio-spinel compounds.\cite{hemberger:05nat} The Cd$^{2+}$ or
Hg$^{2+}$ ions exhibit an electronic configuration of filled
$d^{10}$ shells. Thus on the structural $A$-sites, which form a
diamond lattice, there are no magnetic or orbital degrees of
freedom. The Cr$^{3+}$ ions on the octahedrally coordinated
$B$-sites, which form a pyrochlore lattice, possess half filled
$t_{2g}$ shells and thus also are orbitally inactive. But due to
the quenched orbital moment and Hund's coupling, the $B$-sites
carry a $S=J=3/2$ spin configuration. This simplifies in general
the more complex orbital and magnetic interaction scenarios which
can be found in this structural class of materials, as
illustrated, e.g., by the case of Fe on the $A$-sites.
\cite{buettgen:05njp,fritsch:04prl} The absence of orbital degrees
of freedom in an highly symmetric lattice, changes the
prerequisites for polar order in the compound. Jahn-Teller active
orbital degrees of freedom may trigger non-polar lattice
distortions. Then the cubic lattice relaxes into a Jahn-Teller
distorted, orbitally ordered structure of lower symmetry. These
distortions can be expected to be non-polar due to the symmetry of
the electronic charge distribution of the orbital, to which the
positions of the surrounding ligands are adapted.\cite{hill:00} An
example is the Jahn-Teller active system
FeCr$_2$S$_4$.\cite{fichtl:05prl} In contrast, the absence of any
orbital degrees of freedom, as it is realized via the $d^0$
electronic configuration in canonical ferroelectrics like
BaTiO$3$, allows for the relaxation of the structure into polar
off-center distortions. In \CCS\ and \HCS\ the relatively large
radii of the $A$-site ions destabilize the cubic structure even
more. Accordingly, the tendency towards ionic off-center
distortions is reported in literature for a number of spinel
compounds.\cite{grimes:72pm} In addition, for \CCS\ negative
thermal expansion and the anomalous broadening of x-ray
Bragg-reflections was detected, denoting the softening of the
spinel structure towards lower temperatures.\cite{goebel:76jmmm}
%n addition reports on local structural distortions [L. Hwang et
%al. 1973] and observations of polar order in a number of spinel
%compounds [N. W. Grimes 1972]
Concerning the spin sector, in \CCS\ Cr-spins, located on the
$B$-site pyrochlore lattice, are coupled ferromagnetically,
yielding a transition temperature of $T_c=84$~K. The situation is
more complex in \HCS, which is close to ferromagnetism, but where
the influence of bond-frustration results in the suppression of
the ferromagnetic transition in zero magnetic
field.\cite{tsurkan:06cm} Instead, a spiral-like antiferromagnetic
spin structure is established below $T_N=22$~K. However, the
antiferromagnetic ground state can be overcome already by weak
external magnetic fields and ferromagnetism can be induced. In
addition, even in the absence of an external field, at higher
temperatures evidence for strong ferromagnetic fluctuations was
found.\cite{tsurkan:06cm}

\section{Experimental Details}

The single crystals were grown from the poly-crystalline starting
material by chemical transport with chlorine as transport agent.
The growth experiments were performed at temperatures between 800
and 850~$^\circ$C. Poly-crystalline \HCS\ and \CCS\ samples were
prepared by conventional solid-state reaction in evacuated quartz
ampoules from high purity (99.999\%) binary mercury sulfide and
elementary Cr and S. In order to reach good homogeneity, the
compounds were grinded and re-synthesized several times. The final
tempering was performed at 800~$^\circ$C in an atmosphere of
sulfur excess.  In the following, all data refer to
single-crystalline samples if nothing contrary is explicitly
mentioned.
%The crystals were obtained in form of nearly
%perfect octahedra with shiny surfaces and dimensions up to 2 mm on
%the edge.
All batches were characterized by powder x-ray diffraction and found
to exhibit the normal cubic spinel structure with space group {\it
Fd\=3m} without any indications of impurity phases. The
magnetization measurements were performed with a commercial SQUID
magnetometer (MPMS, {\sc QuantumDesign}) in fields up to 5~T. For
the dielectric measurements, sputtered gold- or silver-paint
electrodes were applied on opposite sides of the plate-like samples.
The complex permittivity (dielectric constant and loss) was
determined over a broad frequency range (3~Hz~$< \nu < 3$~GHz) using
frequency-response analysis and reflectometric techniques.
\cite{schneider:01fe} The electric polarization was detected
directly, employing a high-impedance Sawyer-Tower circuit, or via
integration of the pyro-current measured with an
electrometer.\cite{hemberger:diss} A conventional $^4$He-bath
cryostat and a commercial cryo-magnet (Teslatron, {\sc Oxford})
allowed measurements at temperatures down to 1.5~K and in magnetic
fields up to 10~T.

%1
\begin{figure}[!b]
\begin{center}
    \includegraphics*[width=0.45\textwidth]{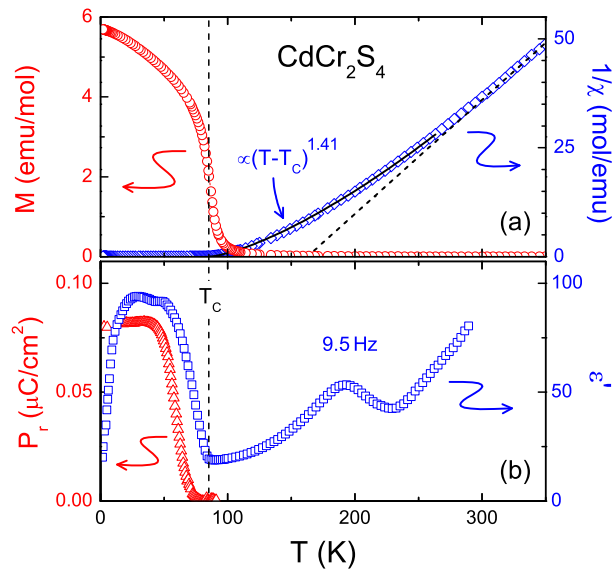}
\end{center}
\caption{Comparison of dielectric and magnetic properties in
\CCS.\cite{hemberger:05nat} (a): Inverse magnetic susceptibility
(right scale) and low field magnetization (left scale) vs.\
temperature measured at field of $H=1$~kOe. (b): Dielectric
permittivity at 9.5~Hz (right scale) and remnant polarization (left
scale) vs.\ temperature. The polarization was obtained via
integration of the pyro-current as measured on heating after poling
in an electric field of 50~kV/m.} \label{1}
\end{figure}

\section{Results and Discussion}

\subsection{\CCS}

%2
\begin{figure}[!bt]
\begin{center}
    \includegraphics*[width=0.35\textwidth]{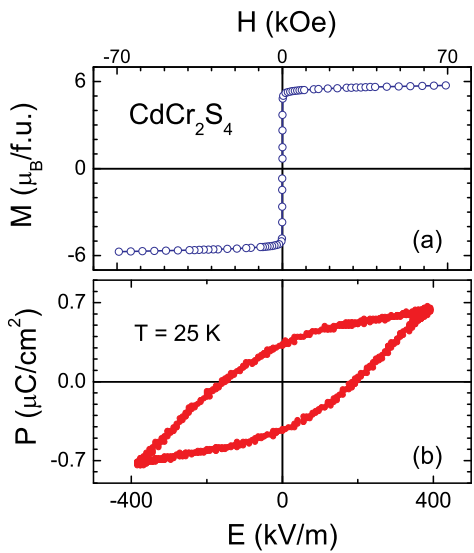}
\end{center}
\caption{Hysteresis loops for the magnetic (a) and dielectric
sector (b) measured at $T=25$~K. The magnetic data measured at
$H>0$ was mirrored and corrected for demagnetization
effects; the polarization data was measured at $\nu=1.13$~Hz. %
(taken from [\onlinecite{hemberger:06sces}]) } \label{2}
\end{figure}

Fig.~\ref{1} compares the dielectric and magnetic behavior of
\CCS. The upper frame (Fig.~\ref{1}a, right scale) shows the
inverse magnetic susceptibility. A typical linear Curie-Weiss type
behavior can be detected at elevated temperatures. The
corresponding fitting values are a Curie-Weiss temperature of
roughly $T_{CW}\approx 150$~K and an effective paramagnetic moment
close to the expected value of $p_{eff}=3.87 \mu_B$ per Cr$^{3+}$.
Already well above the ferromagnetic transition at $T_c\approx
84$~K, clear deviations from the linear behavior are observed due
to the onset of fluctuations. In this regime the inverse
susceptibility can be described as $\chi^{-1}\propto
(T-T_c)^\gamma$ where the exponent $\gamma=1.41$ is close to the
value for the 3d-Heisenberg model of 1.39. Below $T_c\approx 84$~K
spontaneous magnetization sets in (left scale of Fig.~\ref{1})a).
At the same time a steep increase of the dielectric constant can
be detected, which for low frequencies reaches relatively high
values above $\varepsilon'\approx 100$ (Fig.~\ref{1}b, right
scale). The details of the frequency and temperature dependence of
the complex permittivity in the paramagnetic regime will be
discussed below. As shown in Fig.~\ref{1}b (left scale), the low
temperature phase of \CCS\ in addition to spontaneous
magnetization exhibits remnant polarization, which sets in shortly
below $T_c$. Thus the high values of $\varepsilon'$ can be
interpreted as precursor of the onset of remnant dielectric
polarization at lower temperatures. Fig.~\ref{2} shows magnetic
(a) and dielectric (b) hysteresis loops measured at 25~K. $M(H)$
exhibits the characteristics of a soft ferromagnet reaching nearly
the full magnetic moment of $6\mu_B$ expected for two Cr$^{3+}$
ions per formula unit.
%The small
%deviations from the full value can be attributed to pronounced
%magnon features typical for such systems.
The $P(E)$ experiment reveals a well pronounced hysteresis loop.
Complete saturation is not reached within the experimentally
accessible electric field range of 400~kV/m and no well defined
coercive electric field strength can be identified. Such type of
hysteresis behavior is typical for relaxor-ferroelectric behavior,
where nano-scale ferroelectric clusters determine the polarization
response \cite{cross:87fe}.

%3
\begin{figure}[!bt]
\begin{center}
    \includegraphics*[width=0.45\textwidth]{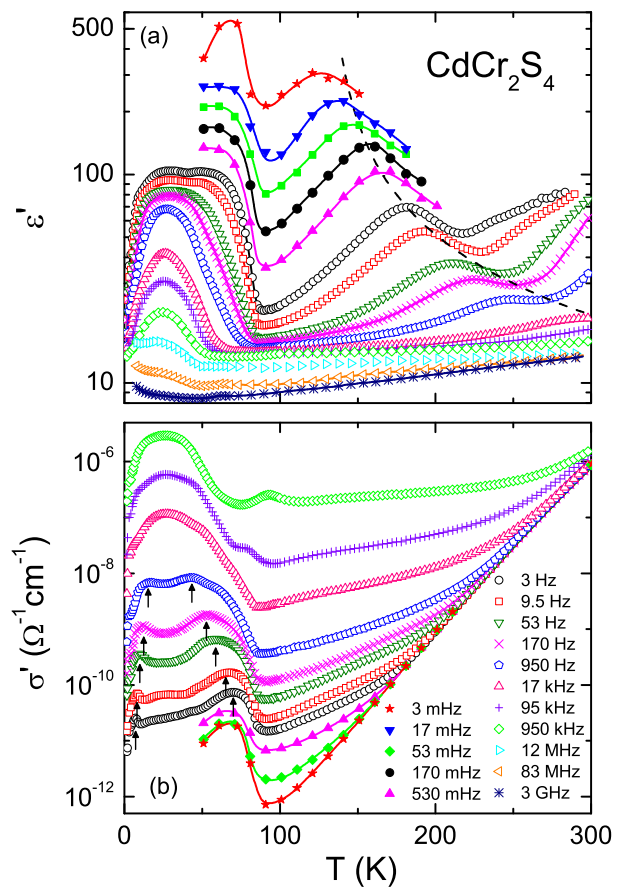}
\end{center}
\caption{Temperature dependence of the real part of the dielectric
permittivity $\varepsilon'$ (upper frame) and the ac-conductivity
$\sigma'\propto \omega \varepsilon''$ (lower frame) of \CCS\ for
various frequencies. The arrows indicate the occurrence of two
maxima in $\sigma'(T)$.  At the magnetic transition $T\approx
84$~K both $\varepsilon'$ and $\sigma'(T)$ exhibit an increase
towards low temperatures ([\onlinecite{lunkenheimer:05prb}];
copyright (2005) by the American Physical Society). } \label{3}
\end{figure}

The relaxor-ferroelectric type of behavior shows up also in the the
frequency and temperature dependence of the complex permittivity.
Fig.~\ref{3} displays dielectric constant $\varepsilon'(T)$ and the
conductivity $\sigma'(T)$ for frequencies from 3~mHz to 3~GHz. In
the paramagnetic regime a peak in $\varepsilon'(T)$ shows up. This
feature shifts to lower temperatures and increases in amplitude with
decreasing frequency. The dashed line indicates a Curie-Weiss law,
$\propto 1/(T-130$~K$)$, for the right flank of the peaks, which can
be taken as an estimate of the static dielectric constant. The
further increase of $\varepsilon'(T)$ towards higher temperatures is
due to contact and conductivity contributions as discussed below.
However, the most characteristic feature in the temperature regime
above 100~K is the described relaxational maximum of the dielectric
constant. This type of behavior in \CCS\ resembles that observed in
relaxor ferroelectrics.\cite{cross:87fe,levstik:98prb} In such
compounds, the reduction of $\varepsilon'$ below the peak
temperature is usually ascribed to a cooperative freezing of
ferroelectric clusters on the time scale given by the frequency of
the applied AC electric-field. This is in contrast to canonical
ferroelectrics where the frequency dependence of $\varepsilon'(T)$
in the sub-GHz regime is negligible, even in the vicinity of the
phase transition. The Arrhenius-type thermal activation of this
relaxation process can be characterized by an energy barrier $E_B
\approx 330$~meV and a high-temperature relaxation time $\tau_0
\approx 1.1\times10^{-11}$~s. The latter value is relatively small
compared to canonical relaxors,\cite{cross:87fe,levstik:98prb} which
may point to a small size of relaxing clusters. The relaxational
features seen in the real part of the permittivity according to the
Kramer-Kronig relation should be accompanied by peaks in the
imagninary part located close to the points of inflection at the
left wing of the maxima in $\varepsilon'(T)$. However, in
Fig.~\ref{3}b, where $\sigma' = \omega \varepsilon_0 \varepsilon''$
($\varepsilon_0$ is the permittivity of the vacuum) is displayed, no
such peaks become obvious. (As the temperature dependence of
$\varepsilon''$ and $\sigma'$ is identical (except for the absolute
values), we chose to plot the latter to avoid a crossing of the
curves and to enhance the readability of the figure.) This is due to
the superimposition by contributions from charge carrier transport.
Above 100~K the data of $\sigma'(T)$, measured at the lowest
displayed frequency of 3~mHz, represents the static DC conductivity.
The levelling-off of the higher frequency data on decreasing
temperature can be understood as AC-conductivity, usually determined
by incoherent charge transport, i.e. hopping processes. The
AC-conductivity usually increases with increasing frequency, which
leads to the staggered behavior of the $\sigma'$ curves for
different frequencies. These additional conductivity contributions
show up in the dielectric loss as $\varepsilon_{AC}'' = \sigma_{AC}
/ (\omega \varepsilon_0)$ and make it difficult to evaluate the
relaxational contributions to the dielectric
loss.\cite{hemberger:05nat,lunkenheimer:05prb}

The most remarkable feature in Fig.~\ref{3} is the strong increase
of $\varepsilon'(T)$ below the ferromagnetic transition
temperature $T_c = 84$~K. It indicates the close coupling of
magnetic and dielectric properties. This behavior can be eyplaine
assuming that the frozen-in dielectric dynamics becomes fast again
at the magnetic transition, thus restoring the large contribution
to the dielectric response of the relaxing
entities.\cite{lunkenheimer:05prb} The dielectric loss monitored
via $\sigma''$ also increases below $T_c$. As mentioned before,
this in part may be influenced by changes in the conductivity.
Magnetoresistive effects have been reported for \CCS\ and
CdCr$_2$Se$_4$ in literature \cite{lehmann:66jap}. However, at
least for the lower frequencies in Fig.~\ref{3}b there is a
two-peak structure, which can be detected within the ferromagnetic
regime. The high-temperature peaks can be interpreted as the loss
maxima connected to the re-acceleration of the relaxation dynmics
below $T_c$. The low temperature peaks mirror the final freezing
at lowest temperatures. The complete temperature dependence of the
intrinsic relaxation time shall be discussed later and is
displayed together with that of \HCS\ in Fig.~\ref{10}.

%4
\begin{figure}[!bt]
\begin{center}
    \includegraphics*[width=0.45\textwidth]{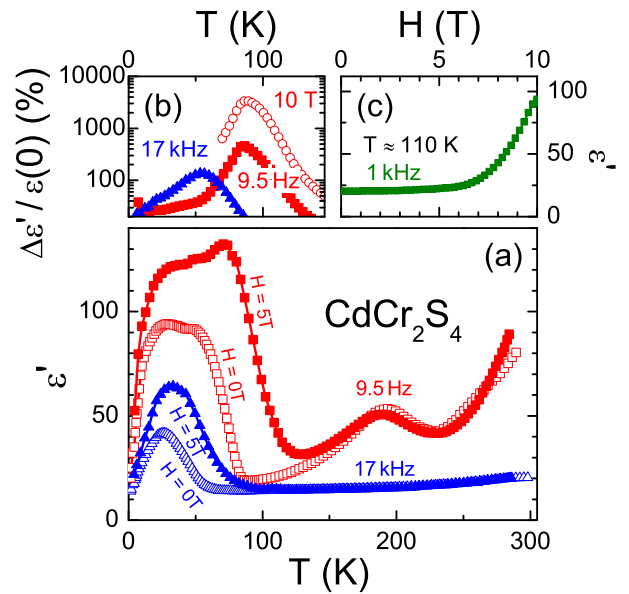}
\end{center}
\caption{Dielectric constant $\varepsilon'(T)$ for frequencies of
9.5~Hz and 17~kHz and magnetic fields of zero and 5~T
(a).\cite{hemberger:05nat,lunkenheimer:05prb} Frame (b): The
temperature dependence of the magneto-permittivity as measured at
5~T (closed symbols) for 9.5~Hz and 17~kHz, and at 10~T ($\circ$)
for 9.5~Hz on a logarithmic scale. In (c) the magnetic field
dependence at $T=110$~K is shown.} \label{4}
\end{figure}

The changes of the relaxation dynamics and the corresponding
increase of the dielectric response below the ferromagnetic
transition are related to the onset of the magnetic order
parameter, the ferromagnetic magnetization, and thus can also be
influenced by an external magnetic field. Fig.~\ref{4}a shows data
of the dielectric constant for two different frequencies in zero
external field and in 5~T. The onset of the rise in
$\varepsilon'(T)$ is shifted to higher temperatures and smeared
out upon the influence of the magnetic field according to the
expected behavior of the magnetization. This results in a strong
magnetic field dependence of $\varepsilon'$ even well above the
spontaneous magnetic transition temperature, which is documented
in Fig.~\ref{4}c. Defining the magneto-capacitance as the ratio of
the magnetic-field induced change of the capacity and the
zero-field value, one arrives at values of up to 500~\% in a field
of 5~T and up to 3000~\% in a field of
10~T\cite{lunkenheimer:05prb} as displayed in in Fig.~\ref{4}b. In
analogy to the similar scenario of a magnetic-field induced shift
of a magnetic transition, causing the occurrence of colossal
magneto-resisitive effects (CMR), e.g. in manganites, one may note
the above described behavior in \CCS\ as colossal
magneto-capacitive effect (CMC). However, this effect appears to
be strongly frequency dependent as it seems to be intimately
coupled to the dynamics of the ferroelectric relaxor entities
dominating the dielectric response.

\subsection{Doped and poly-crystalline \CCS}

%5
\begin{figure}[!bt]
\begin{center}
    \includegraphics*[width=0.45\textwidth]{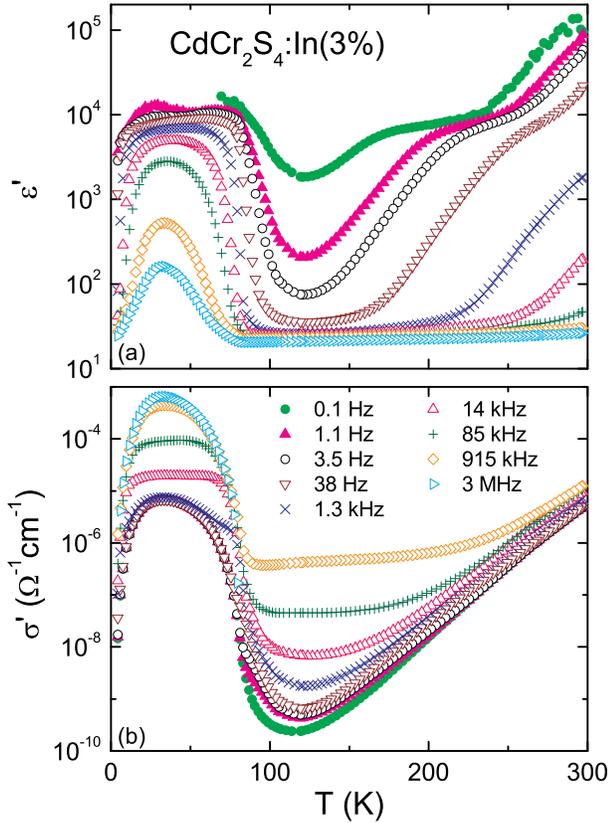}
\end{center}
\caption{Temperature dependence of the real part of the dielectric
permittivity $\varepsilon'$ (upper frame) and the ac-conductivity
$\sigma'\propto \omega \varepsilon''$ (lower frame) of \CCS\ doped
with 3\% In for various frequencies. } \label{5}
\end{figure}

%5b
%\begin{figure}[!bt]
%\begin{center}
%    \includegraphics*[width=0.45\textwidth]{5b-CCISe-Tdep.eps}
%\end{center}
%\caption{Temperature dependence of real part of the dielectric
%permittivity $\varepsilon'$ (upper frame) and the ac-conductivity
%$\sigma'=\omega \varepsilon''$ (lower frame) of \CCS:Se(5\%) for
%various frequencies. } \label{5b}
%\end{figure}

%6
\begin{figure}[!bt]
\begin{center}
    \includegraphics*[width=0.45\textwidth]{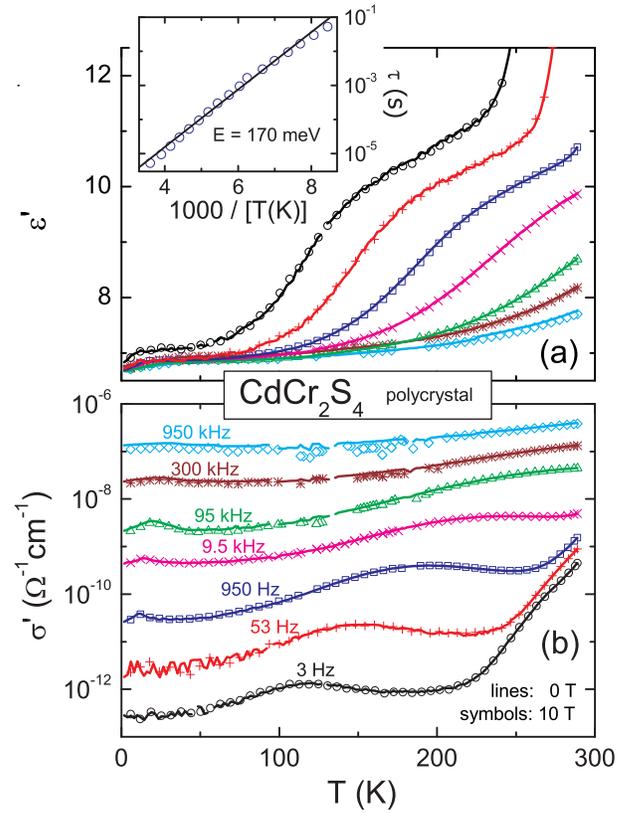}
\end{center}
\caption{Temperature dependence of the real part of the dielectric
permittivity $\varepsilon'$ (upper frame) and the ac-conductivity
$\sigma'\propto\omega \varepsilon''$ (lower frame) of
poly-crystalline \CCS\ for various frequencies. } \label{6}
\end{figure}

The results shown so far have been obtained from samples of
single-crystalline \CCS. In the following, we want to address the
sample dependence of the dielectric properties by providing
results on doped and poly-crystalline material. Fig.~\ref{5}
displays the real parts of the dielectric permittivity and
AC-conductivity of \CCS\ doped with 3~\% In. The In-doping is
supposed to create In$^{3+}$ defect states and to introduce
additional charge carriers to the system. Correspondingly the
conductivity as shown in Fig.~\ref{5}b at room temperature is
enhanced by about one decade compared to the nominally undoped
\CCS\ (Fig.~\ref{3}). However,
%already at this point it shall be  mentioned that
due to the use of chlorine as transport agent for the crystal
growth, even in the pure compound a finite charge doping in the
sub-percent range has to be considered. The general behavior of the
In-doped compound is to be similar to that of the undoped one. Below
the magnetic transition at roughly 80~K a pronounced increase of
both quantities, $\varepsilon'$ and $\sigma'$, is detected. In that
temperature range the shape of the broad maximum in the conductivity
even displays signs of the discussed two-peak structure, compatible
to the re-accelerating and (towards lowest temperatures) re-freezing
of the relaxor dynamics. However, the leveling-off in $\sigma'$
above the magnetic transition temperature sets in only for the
highest frequencies. In the MHz-range similar values of about
$10^{-6}/\Omega$cm are measured for the pure and the In-doped
compound. This reflects a similar contribution from hopping
transport mechanisms showing up in the AC-conductivity. But in
\CCS:In(3\%) the DC-transport is much more dominating. This hampers
the ability to measure dielectric polarization or pyro-currents and
to distinguish them from ohmic contributions. As documented in
Fig.~\ref{5}a, the absolute values of the dielectric constant are
strongly enhanced compared to pure \CCS, reaching step heights up to
$10^4$. At this point it is important to note that the
DC-conductivity also exhibits strong changes at the magnetic
transition (base on the fact that the $\sigma(T)$ curves for the
lowest frequencies in Fig.~\ref{5}b coincide below 80~K, the
DC-conductivity in this region is well approximated by the curve at
0.1~Hz).
However, $\sigma_{DC}$ does of cause not have any influence on the
dielectric constant. This is in contrast to the discussed
AC-contributions of the conductivity which via the Kramers-Kronig
relation has to appear in $\sigma'$ and $\sigma''$ and thus also
leads to a contribution to the dielectric constant $\varepsilon'
\propto \sigma''/\omega$.\cite{jonscher:83} However, the steps in
the dielectric constant, discussed already for pure \CCS\ in the
paramagnetic regime still can be detected. The relaxational
character of this contribution, together with the fact that the
AC-conductivity seems to be of similar magnitude for the doped and
the undoped compound, points against an origin related to incoherent
transport processes. The thermal activation of this relaxational
process in \CCS:In(3\%) can be characterized by $E_B \approx
200$~meV and $\tau_0 \approx 10^{-7}$~s. The latter value is
considerably higher than in nominally undoped \CCS. This could be
interpreted with larger relaxing clusters, which also would be
compatible with the enhanced values of the permittivity.

However, towards higher temperatures an increasing background, like
in \CCS, comes into play. This contribution may indeed arise from
non-intrinsic, so-called Maxwell-Wagner type,
contributions.\cite{wagner:13ap,lunkenheimer:04prb} At the sample
surfaces the generation of a Schottky-type of diode between the
metallic electrode and the semiconducting bulk material can give
rise to a depletion layer, possessing high resistance and
capacitance. Both can be modeled by a parallel RC-element connected
in series with the bulk sample, leading to a characteristic
relaxation time, which changes due to the temperature dependence of
the conductivity. In this case the observed relaxation feature would
not be a bulk property. Prominent examples for Maxwell-Wagner
relaxations with very high (typically $10^3$-$10^5$), non-intrinsic
values for $\varepsilon'$ are the so-called "colossal dielectric
constant" materials \cite{lunkenheimer:04prb}. Such effects may
additionally influence the increase of $\varepsilon'(T)$ above the
discussed intrinsic relaxation step. A possibility to distinguish
between both scenarios is the variation of the electrode material by
using contacts made from silver-paint or sputtered gold. In both
cases discussed so far, namely nominally pure single-crystalline
\CCS\ (Fig.~\ref{3}) and \CCS:In(3\%) (Fig.~\ref{5}) significant
differences between the results obtained for different electrode
materials could only be detected for the non-intrinsic
high-temperature wing of the permittivity, showing up, e.g., in
Fig.~\ref{5}a at $T>240$~K at $\nu=0.1$~Hz. In contrast, as
demonstrated for the pure compound in
[\onlinecite{hemberger:06sces}], the relaxation features discussed
above are still observed at the same temperatures with the same
amplitude.

An alternative approach is to regard stoichiometric but
poly-crystalline \CCS.  Fig.~\ref{6} represents plots of
$\varepsilon'(T)$ and $\sigma'(T)$ for pressed powder samples. Here,
in contrast to the \CCS\ single crystals, traces of Cl and
corresponding impurity charge doping can not be excludd.
Accordingly, the conductivity is much reduced as displayed in
Fig.~\ref{6}b. The complete temperature range is dominated by
incoherent AC-conductivity, and only for the lowest measured
frequencies, in the Hz-range, the characteristic increase with
increasing temperature typical for thermally activated DC-transport,
can be observed.
Compared to the single crystalline sample (Fig.~\ref{3}b), the
AC-conductivity is reduced by less than one decade only, while,
e.g.\ at room temperature, the DC-conductivity is about three
decades smaller in the poly-crystal.
Also the dielectric response in the poly-crystal is much weaker.
Nevertheless, a relaxational step in $\varepsilon'(T)$ for
temperatures above 100~K can be observed. The thermal activation of
this relaxation process (see inset of Fig.~\ref{6}) in
poly-crystalline \CCS\ can roughly be characterized with an
activation barrier of about $E_B \approx 170$~meV and $\tau_0
\approx 6\times 10^{-9}$~s, both values being compatible with the
parameter range of the single crystals discussed above. Thus this
much smaller dielectric contribution could be interpreted as a
remainder of the pronounced single-crystal phenomena. However, the
strong anomaly at the magnetic transition is completely absent in
the poly-crystal. In addition, no magnetic field dependence can be
found in the vicinity of the magnetic transition, which still occurs
at about 84~K (compare the solid lines (0~T) and the symbols (10~T)
in Fig.~\ref{6}). This seems to point to the necessity of additional
charge carriers and/or defect states for the microscopic coupling
mechanism between dielectric and magnetic polarization. On the other
hand, the absence of inner strains and the possible variation of
stoichiometry at the grain boundaries could play an important role
for the suppression of the magneto-capacitive effects in the
poly-crystalline material. In this context it is interesting that
annealing of the single-crystalline \CCS\ samples, both, in vacuum
or sulphur atmosphere, leads to a suppression of the frequency
dependent  relaxation features above and below $T_c$ and no remanent
electric polarization can be found at low temperatures. Further
systematic studies of this issue are highly needed and currently
being performed.

\subsection{\HCS}

%7a
\begin{figure}[!bt]
\begin{center}
    \includegraphics*[width=0.45\textwidth]{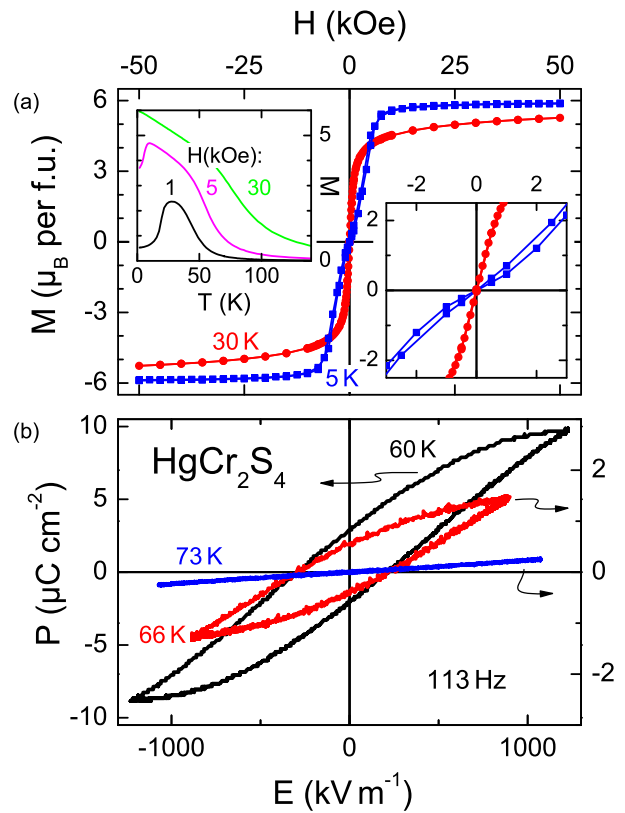}
\end{center}
\caption{Hysteresis loops for the magnetic ($M(H)$, (a)) and
dielectric sector ($P(E)$, (b)) measured at various temperatures for
\HCS. The polarization data was measured at $\nu=113$~Hz
([\onlinecite{weber:06prl}]; copyright (2005) by the APS). The
insets show the temperature dependence of the magnetization for
various fields and a magnification of the magnetic hysteresis loops
for low fields.} \label{7a}
\end{figure}

%7b
\begin{figure}[!bt]
\begin{center}
    \includegraphics*[width=0.45\textwidth]{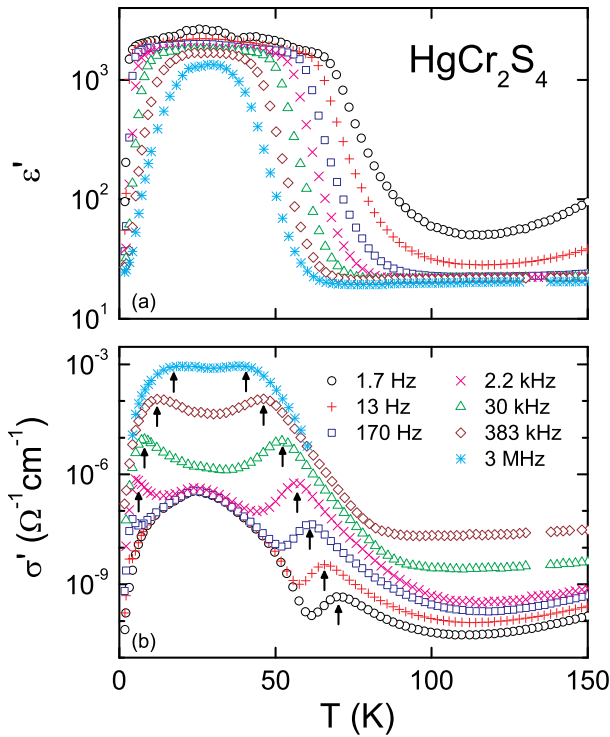}
\end{center}
\caption{Temperature dependence of real part of the dielectric
permittivity $\varepsilon'$ (a) and the ac-conductivity
$\sigma'\propto\omega \varepsilon''$ (b) of \HCS\ for various
frequencies ([\onlinecite{weber:06prl}]; copyright (2005) by the
APS).} \label{7b}
\end{figure}

Even though the observed magneto-capacitive effects are relatively
sensitive to the details of the sample preparation and impurity
concentration, they seem to be stable against the exchange of the
chalcogen ion. Replacing sulphur by selenium leads to
CdCr$_2$Se$_4$, in which the ferromagnetic transition is increased
towards $T_c \approx 125$~K.\cite{hemberger:06sces} At this
temperature a pronounced variation of $\varepsilon'(T)$ can be found
denoting a similarly strong magneto-dielectric coupling like in
\CCS. Also the typical relaxor peaks in the real part of the
permittivity at $T
> T_{c}$ are shifted to higher temperatures in CdCr$_2$Se$_4$. In
the following we discuss the compound \HCS\ where instead of the
chalcogen the $A$-site ion is replaced. The different ionic size in
the first instance changes the bond-distances of the spinel
structure, which, nevertheless, stays cubic at room
temperature.\cite{tsurkan:06cm} This induces a redistribution of the
interaction strengths between nearest and next-nearest
super-exchange as well as direct $B$-$B$ exchange pathes. This
results in a bond-frustrated scenario with complex magnetic
properties. Below $T_N=22$~K a helical antiferromagnetic structure
is established in the absence of a magnetic field. However, already
in small external magnetic fields ferromagnetic order can be
induced. Fig.~\ref{7a}a shows magnetic hysteresis measurements for
$T=5$ and 30~K. The slight S-shape of the low-temperature curve,
observed at small fields, denotes the metamagnetic transition from
antiferromagnetic into induced ferromagnetic order reaching the full
expected ordered moment of 6~$\mu_B$ (3~$\mu_B$ per Cr$^{3+}$) in
fields above 1~T. Even though at temperatures above $T_N$ no
spontaneous ferromagnetic order can be detected (as monitored in the
upper inset of Fig.~\ref{7a} via the $M(T)$ curve measured at 1~kOe)
in fields even below 1~T a large ferromagnetic component can be
induced. This denotes the presence of strong ferromagnetic
fluctuations and the vicinity of the system to ferromagnetic
order.\cite{lehmann:70prb} At the same time (Fig.~\ref{7a}b) a
dielectric polarization hysteresis appears for temperatures below
70~K. The values for the remanent polarization in \HCS\ are even
higher than in \CCS. Due to the finite conductivity in \HCS\ (shown
below), the interpretation of such type of dielectric hysteresis
measurements is not completely unambiguous. In
ref.~\onlinecite{pintilie:05apl} possible reasons for non-intrinsic
non-linearities of the dielectric response are discussed. One
indication for extrinsic behavior could be the frequency dependence,
which is present in the case of \HCS. This criterion is, however, of
course also compatible with the picture of a relaxor ferrolectric
state. A second criterion would be the vanishing of a saturation for
increasing frequencies, which is in contrast to the experimental
findings.\cite{weber:06prl} An alternative determination of the
polarization via measurements of the pyro-current unfortunately
fails due to the relative high conductivity of the samples. In
addition, the variation of the electrode material and the
investigation of different sample batches yield similar dielectric
properties, which points against an extrinsic nature of the observed
phenomena.\cite{weber:06prl}

%8
\begin{figure}[!bt]
\begin{center}
    \includegraphics*[width=0.35\textwidth]{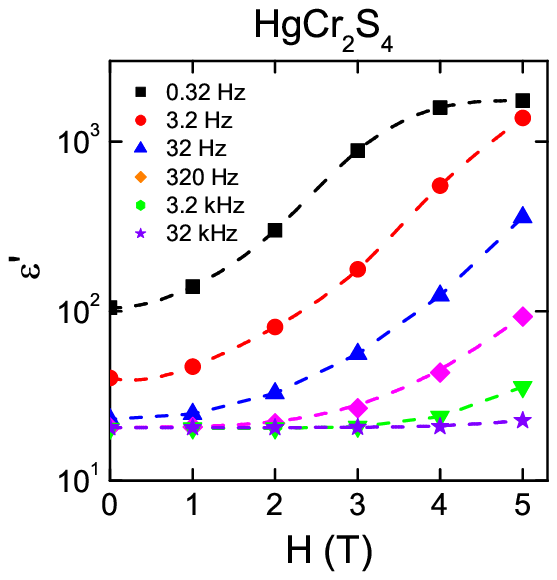}
\end{center}
\caption{Magnetic field dependence of $\varepsilon'$ in \HCS\ for
various frequencies at $T=120~K$.} \label{8}
\end{figure}

%9
\begin{figure}[!bt]
\begin{center}
    \includegraphics[width=0.45\textwidth]{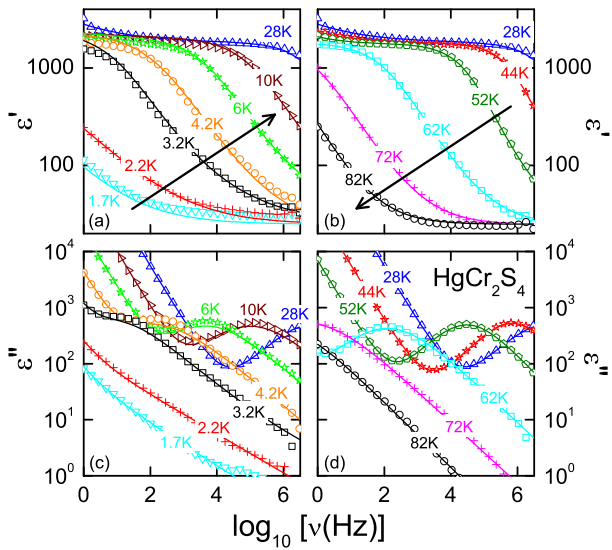}
\end{center}
\caption{Frequency dependence of the complex dielectric permittivity
$\varepsilon^*$ ((a,b): real part; (c,d): loss) as measured for
various temperatures in \HCS. The arrows indicate the shift of the
mean relaxation rate $\nu_0 = 1/2\pi\tau_0$ with temperature.}
\label{9}
\end{figure}

Fig.~\ref{7b} displays the temperature dependence of the real parts
of dielectric constant and conductivity for various frequencies in
the temperature range below 150~K. The response at higher
temperatures is dominated by extrinsic Maxwell-Wagner type
contributions as indicated by the dependency of the the data on the
choice of electrode material.\cite{weber:06prl} The results of
Fig.~\ref{7b} qualitatively resemble those of the findings in \CCS\
and \CCS:In(3\%) in the corresponding temperature regime. The
conductivity is significantly higher than in \CCS.
%  \HCS\ $E = 280$~meV, $tau_0 = 4\times 10^{-9}$~s
A double peak structure in $\sigma'(T)$ at $T<75$~K can clearly be
resolved. This behavior again can be explained by the
re-acceleration of relaxor entities due to the presence of strong
ferromagnetic correlations and the subsequent re-freezing towards
lower temperatures. The peaks in the conductivity coincide with the
points of inflection in the dielectric constant (Fig.~\ref{7b}a).
The plateau-like maximum of $\varepsilon'$ reaches a value of
roughly 2000. The increase of the dielectric constant for
temperatures between 40~K (high frequencies, 3 MHz) and about 70~K
(low frequencies, 1.7~Hz) is much stronger than in \CCS.
Correspondingly, a pronounced magnetic field dependence of the
permittivity can be expected in the regarded temperatures range.
Fig.~\ref{8} displays the dielectric constant $\varepsilon'(H)$ in
magnetic fields up to 5~T for various frequencies. The application
of an external magnetic field enhances the ferromagnetic
correlations and shifts their appearance towards higher
temperatures. As a consequence the mentioned re-acceleration of the
dielectric dynamics is induced by the magnetic field and
$\varepsilon'$ considerably increases with increasing field. This
effect is strongly frequency dependent again underlining the
relaxational character of the magneto-capacitive coupling. At a
temperature of 120~K and for frequencies in the Hz-regime a magnetic
field of 1~T easily can change the permittivity by a factor of ten.
Under the same conditions the response for frequencies above 30~kHz
is barely effected. The details of the frequency dependence of the
complex permittivity are monitored in Fig.~\ref{9}. The
representation as permittivity spectra $\varepsilon^*(\omega)$
contains the most significant information on the relaxation
dynamics. The condition $\tau=1/2\pi\nu$ where $\nu$ is the position
of the maximum in the dielectric loss or the point of inflection in
the step of the dielectric constant allows for the straightforward
determination of the relaxation time $\tau$. Typical relaxation
steps in $\varepsilon'(T)$ are monitored for various temperatures in
Figs.~\ref{9}a and b, and the accompanying loss peaks are observed
in $\varepsilon''(T)$ in Figs.~\ref{9}c and d. (Note that the
increase in the loss spectra observed towards low frequencies
results from finite DC-conductivity as discussed below). Considering
the steps in $\varepsilon'$ in the low temperature region
(Fig.~\ref{9}a), the usual thermally activated scenario becomes
apparent: They shift to higher frequency, i.e. the dynamics becomes
faster with increasing temperature. But in the temperature range
around the onset of ferromagnetic fluctuations between 30 and 80~K
(Fig.~\ref{9}b) the opposite behavior is found: The system seems to
slow down with increasing temperature! Obviously, in this regime not
the temperature but the magnetic correlations seem to determine the
relaxation rate. On the other hand the DC-conductivity of \HCS\
experiences very pronounced changes at about 80~K and also exhibit a
strong magnetic field dependence.\cite{lehmann:66jap, weber:06prl}
Thus this material exhibits CMR and CMC properties at the same time.
However, a direct relation between both quantities is not obvious,
as has been pointed out above, and the relaxational character of the
dielectric response does not point towards charge transport as the
origin of the strong anomalies in $\varepsilon'$. Strong
spin-disorder scattering has been postulated for ferromagnetic
semiconductors with very small charge carrier
concentrations.\cite{majumdar:98nat} An enhanced local electronic
polarizability could link the charge transport to the relaxor
ferroelectric properties in these slightly doped spinel systems. But
further investigations and a more detailed theoretical approach
would be necessary to settle such a picture.

%10
\begin{figure}[!bt]
\begin{center}
    \includegraphics*[width=0.4\textwidth]{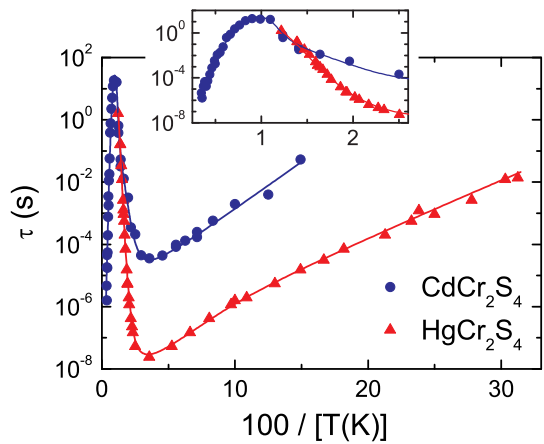}
\end{center}
\caption{Comparison of the temperature dependent mean relaxation
times of \HCS\ and \CCS\ in an Arrhenius-type presentation. The
inset displays a magnification of the high temperature behavior.}
\label{10}
\end{figure}

A quantitative evaluation of the results shown in Fig.~\ref{9} was
obtained from the simultaneous fitting of the real and imaginary
part of the permittivity data, the resulting fit curves being
displayed as solid lines. The fits account for an symmetric
(Cole-Cole) distribution of relaxation times centered at
$\tau_0(T)$, DC-conductivity and DC-dielectric constant, and an
AC-conductivity contribution $\sigma_{AC}'\propto\nu^s$ with
$s<1$.\cite{jonscher:83} The latter term via the Kramers-Kronig
relation is related to the imaginary part of the conductivity and
thus also gives a contribution to the real part of the permittivity
$\varepsilon'\propto \sigma''/\omega$. In literature this power-law
type of contribution is referred to as "universal dielectric
response" and usually assumed to account for hopping transport
processes.\cite{jonscher:83} The mean relaxation times resulting
from this type of evaluation are displayed as $\tau(1/T)$ in
Fig.~\ref{10} for \HCS\ and \CCS. At high temperatures (small values
of $1/T$; see inset of Fig.~\ref{10}) the mentioned conventional
relaxational freezing scenario, corresponding to an increase of
$\tau$ with decreasing $T$, can be followed for \CCS. Then for both
compounds an anomalous regime occurs, in which the relaxation times
become reduced for decreasing temperatures. In this regime the
magnetic field and the magnetization lead to a re-acceleration of
the dynamics and the frozen relaxor-system melts again. Finally, at
lowest temperatures, where the influence of the magnetization
saturates, a scenario of thermal activation becomes dominant again
and the dynamics slow down again. In this regime the relaxation time
in \HCS\ is about two decades lower than in \CCS. This difference is
difficult to explain considering the different magnetic
configuration of both compounds. As ferromagnetism seems to decrease
$\tau$, the completely ferromagnetically ordered \CCS\ should rather
show the faster dielectric response. The in general larger
dielectric response in \HCS\ also would point towards larger relaxor
entities with accordingly higher relaxation times.

\section{Conclusions and Summary}

The present dielectric experiments provide clear evidence for an
relaxational process using the observed magneto-capacitive
phenomena.
%A scenario in which this relaxation mechanism interacts
%with magnetic order via exchange-striction can be considered the
%most plausible.
The ferroelectric distortions are supposed to result from an
off-center position of the Cr$^{3+}$-ions which generates a locally
polar but macroscopically isotropic cluster
state.\cite{grimes:72pm,hemberger:05nat} Especially in \HCS\ bond
frustration of the magnetic sector may play a considerable role. But
the structural frustration within the cubic lattice, which at room
temperature is highly symmetric and geometrically unconstrained, is
supposed to be the constituting ingredient driving the observed
relaxor-like freezing.\cite{ramirez:03nat} However, recent LDA+U
calculations did not find any indication for the softening of a
polar phonon mode which could explain the local ferroelectric
distortions.\cite{fennie:05prb} Nevertheless, pronounced spin-phonon
coupling could be investigated via optical spectroscopy in magnetic
fields and will be reported in a forthcoming paper.
%Also from high-resolution sc x-ray diffraction
% structural anomalies (Krimmel priv. comm.) ?!
On the other hand, the details of the coupling mechanism between
magnetization or the magnetic field and the dielectric permittivity
in these compounds is so far unclear. As pointed out above, the
relaxation dynamics of the polar moments is accelerated below $T_c$,
but the microscopic origin of the detected relaxation dynamics and
why this dynamics couples so strongly to the magnetic order
parameter is still unknown. Most plausible is a coupling via
exchange striction, i.e. volume changes arising from the magnetic
exchange energy.\cite{martin:69jap} The onset of spin order leads to
a local magnetostrictive distortion of the lattice. Thus the energy
barriers against dipolar reorientation are reduced and the mean
relaxation rate is enhanced.
 An alternative explanation could be a
magnetic-field induced variation of charge-carrier mobility or
density. As mentioned before, a sizable DC magneto-resistive effect
is well known for \CCS\ and \HCS,\cite{lehmann:66jap,weber:06prl}
Even so the DC-conductivity of course not directly contributes to
$\varepsilon'$, but only to $\varepsilon''$. Its variation may
denote changes in the electronic
polarizability.\cite{monceau:01prl,staresinic:06prl} Within such a
scenario one could expect a strong enhancement of the dielectric
constant in the vicinity of a metal-insulator transition, e.g.
driven by charge doping of
semiconductors.\cite{castner:80,aebischer:01prl} However, the
influence of purely electronic degrees of freedom probably can not
explain the observed dispersion in the investigated frequency range.
Finally, even though the influence of the electrode polarization via
Maxwell-Wagner type of processes appears to be unlikely if
considering our experiments with different types of contacts, the
importance of possible internal boundary mechanisms can not be
completely ruled out, and needs further clarification.\cite{toda:72}

In summary, our results reveal that in \CCS\ and \HCS\ ferromagnetic
order, or at least strong ferromagnetic fluctuations, coexists with
a relaxor-ferroelectric state, characterized by a significant
relaxational behavior. Both properties are strongly coupled. The
observed strong increase of the dielectric permittivity in the
magnetically ordered state of these compounds can be explained by a
re-acceleration of the dielectric relaxor dynamics. These are
influenced by the onset of magnetization, which can be driven by
external magnetic fields. This results in a colossal
magneto-capacitive effect, which exhibits a pronounced frequency
dependence. Concerning the microscopic origin of the observed
effects further investigations are needed.
% on the microscopic origin of this puzzling behavior.

%\cite{tristan:05prb}

\acknowledgments

This work was partly supported by the Bundesministerium f\"ur
Bildung und Forschung (BMBF) via Grant No. VDI/EKM 13N6917-A and
by the Deutsche Forschungsgemeinschaft via Sonderforschungsbereich
SFB 484 (Augsburg).

%\bibliographystyle{apsrev}
%\bibliographystyle{prsty}
%\bibliography{jabbr,joalt,joneu}

\end{document}